\documentclass[aps,prd,twocolumn,superscriptaddress,floatfix,showpacs]{revtex4-1}

\usepackage{amsmath}
\usepackage{amsfonts}
\usepackage{bm}
\usepackage{graphicx}

\newcommand{\mM}{\mathcal{M}}
\newcommand{\Msun}{M_{\odot}}

\begin{document}


\title{Supermassive Black Hole Tests of General Relativity with eLISA}


\author{C\'{e}dric \surname{Huwyler}}
\email[]{chuwyler@physik.uzh.ch}
\affiliation{Physik-Institut, Universit\"at Z\"urich,
Winterthurerstrasse 190, 8057 Z\"urich, Switzerland}

\author{Edward K. \surname{Porter}}
\affiliation{Fran\c{c}ois Arago Centre, APC, Universit\'e Paris Diderot,\\ CNRS/IN2P3, CEA/Irfu, Observatoire de Paris, Sorbonne Paris Cit\'e, \\10 rue A. Domon et L. Duquet, 75205 Paris Cedex 13, France}

\author{Philippe \surname{Jetzer}}
\affiliation{Physik-Institut, Universit\"at Z\"urich,
Winterthurerstrasse 190, 8057 Z\"urich, Switzerland}


\date{\today}

\begin{abstract}
Motivated by the parameterized post-Einsteinian (ppE) scheme devised by Yunes
and Pretorius, which introduces corrections to the post-Newtonian coefficients
of the frequency domain gravitational waveform in order to emulate alternative 
theories of gravity, we compute analytical time domain waveforms that, after a 
numerical Fourier transform, aim to represent (phase corrected only) ppE 
waveforms.  In this formalism, alternative theories manifest themselves via corrections to the phase and frequency, as predicted by
General Relativity (GR), at different post-Newtonian (PN) orders.   In order to present a generic test of alternative theories of gravity, we
assume that the coupling constant of each alternative theory is manifestly positive, allowing corrections to the GR waveforms to be
either positive or negative.  By exploring the capabilities of massive black hole binary GR waveforms in the detection 
and parameter estimation of corrected time domain ppE signals, using the current eLISA configuration (as presented for the ESA Cosmic Vision L3 mission),
we demonstrate that for corrections arising at higher than 1PN order in phase and frequency, GR waveforms are sufficient for both detecting and estimating the parameters
of alternative theory signals.  However, for theories introducing corrections at the 0 and 0.5 PN order, GR waveforms are not capable of covering
the entire parameter space, requiring the use of non-GR waveforms for detection and parameter estimation.
\end{abstract}


\maketitle

\section{Introduction}

General Relativity (GR) has been tested rigorously in the recent past \cite{will2014}; so far, no evidence has been observed in the macroscopic regime that suggests any failure of GR.
Nevertheless, various alternative theories of gravity have been proposed in order to account for effects that are currently otherwise explained, or for the lack of a common
intersection between GR and quantum field theory. A few of these theories can be ruled out by solar system and binary pulsar observations.  However,  many of them are still essentially
unconstrained since GR has never been tested in the true strong field regime where $v/c$ approaches unity or where $\Phi/c^2 = GM/rc^2\sim1$, where $G$ is Newton's gravitational constant, $c$ is the speed of light, $r$ is the effective distance of measurement and $M$ is the mass of the system.

Gravitational waves (GWs) will provide a unique opportunity to test GR in the strong field, dynamical regime. A ground-based network of detectors (i.e.\ Advanced LIGO, Advanced Virgo, KAGRA etc.) is currently
being enhanced, with the aim of going online in 2015 and providing the first direct detection of astrophysical sources of GWs in the Hz-kHz regime within the next decade. 
Simultaneously, pulsar timing array analysis is expected to improve to a point where detection in the nHz regime should be possible within the same time frame.

On a longer time scale, ESA has recently chosen the theme of the \textquotedblleft Gravitational Wave Universe\textquotedblright for the ESA Cosmic Vision L3 mission in order to nourish the development of a space-based GW mission. The mission, called eLISA,  consists of a triangle of three spacecrafts in a heliocentric orbit, interconnected with two laser arms.  This single channel laser interferometer will operate at frequencies between $\sim10^{-5} - 1$ Hz, and  will be sensitive to supermassive black hole binaries (SMBHBs) with total redshifted masses between $10^4-10^8 \Msun$. SMBHBs provide excellent strong-field tests of GR, as the waveform models for such objects are 
quite well understood \cite{blanchet2014}.  The waveforms are composed of three phases : inspiral, merger and ringdown. The inspiral phase is adequately described by post-Newtonian theory, while the ringdown phase is known from 
perturbation theory.  Incredible advances in numerical relativity now mean that we have a much more complete idea of how the merger phase works.  In the next
few years as this field improves even further, we will rapidly approach a point where we may have complete analytical waveforms involving all three phases.

Tests of GR can be performed from a number of different viewpoints \cite{yunessiemens2013}. 
For \emph{direct} tests, one takes a certain alternative model to GR with known action, e.g. a scalar-tensor theory, computes the modified gravitational
waveforms and checks (in a post detection manner) whether or not they achieve a higher correlation with the recorded data than GR waveforms. 
The advantage of such a top-down approach is that it is possible to directly constrain the coupling constant(s) of the theory through evaluating the detector data. 
At the same time, there is of course the disadvantage that GR can only be tested against this specific theory.

Since it requires an intense effort to perform such steps for every imaginable alternative to GR, one can think of performing more \emph{generic} tests.
Certain features of GR can be tested in this more phenomenological manner: what if the `graviton' had a mass? What if Lorentz invariance was violated? What if Newton's gravitational
constant changed with time? 
Such features could be exhibited by a certain subset of alternative theories; and while a generic test will not be able to reveal which particular alternative
is the true underlying theory, it could certainly provide evidence against GR and point us in the right direction.

As a simple example, consider a class of theories that exhibit massive gravity. There have been several studies
assessing the ability of ground and space-based gravitational-wave detectors to see whether such an effect is manifested in the detector data 
\cite{will1998, berti2005, stavridiswill2009, arunwill2009, stavridiswill2009, keppelajith2010, yagitanaka2010, bertietal2011, delpozzoveitch2011, huwyler2012}.
Will \cite{will1998} has computed the effect of a GW dispersion relation through the different arrival times of wave trains with different frequencies 
(to leading order) as a 1 post-Newtonian (PN) correction to the frequency domain GR phase, namely:
\begin{align}
 \Psi_{\text{MG}}(f) &= \Psi_{\text{GR}} - \beta_{\text{MG}}\, u^{-1}, \\
 \beta_{\text{MG}} &= \frac{\pi^2 D(z) \, G {\mathcal M}}{c^2 \lambda_g^2(1+z)},
\end{align}
where $f$ is the GW frequency, $u = \frac{G \mM}{c^3} \pi f$ is the reduced GW frequency, and ${\mathcal M} = m\eta^{3/5}$
is the chirp mass of the binary.  In this last expression $m=m_1+m_2$ and $\eta=m_1 m_2 / m^2$ are the total mass and the symmetric mass ratio of the source.
Finally,  $z$ is the redshift of the source, and the distance parameter is given by $D(z) = \frac{1+z}{a_0} \int_{t_e}^{t_a} a(t) \, dt$,
where $t_e$ and $t_a$ are the times of signal emission and arrival, respectively, and $a(t)$ is the cosmic scale factor with present value $a_0=a(t_a)$.
$\beta_{\text{MG}}$ is in functional relation to the coupling constant of the massive gravity correction, the gravition's Compton wavelength $\lambda_g$. From an analysis perspective,
in the remainder of this paper we call $\beta_{\text{MG}}$ the coupling constant of this particular physical effect.

Being interested in whether or not GR is the correct theory, it is reasonable to perform as many such tests as possible, without having to assume a certain alternative theory
or a particular physical effect. This can be done with a waveform model that aims to catch any possible deformation of the expected GR waveform. Among such are tests that 
introduce perturbations to the PN coefficients of the GR inspiral waveform \cite{arunetal2006_1,yunespretorius2009,lipozzo2012}; similarly, the ringdown part 
of the waveform can be checked \cite{dreyer2004,bertietal2006,yunespretorius2009,veitchetal2012}.
One has to be careful here, as one de-facto tests only the PN coefficients of GR or the perturbed Kerr metric, respectively, but not directly GR itself.
Although it has been shown that phenomenological inspiral-merger-ringdown (IMR)
waveforms can in principle decrease parameter estimation errors by almost an order of magnitude \cite{keppelajith2010}, we consider only the inspiral part in this work due to the lack of
a full theoretical model for the merger phase. A framework to test the PN coefficients of the inspiral waveform that has been studied recently is the parameterized post-Einsteinian (ppE) 
scheme devised by Yunes and Pretorius \cite{yunespretorius2009, cornishsampson2011, sampsonyunescornish2013,sampsoncornishyunes2014} which is motivated by the stationary 
phase approximation (SPA) and works in the frequency domain. There, leading order corrections to the amplitude and the phase of the waveform are introduced:
\begin{equation}
 \label{eq:ppE}
 \tilde{h}_\text{ppE}(f) = \tilde{h}_\text{GR}(f) \left( 1 + \alpha u^a \right) e^{i\beta u^b},
\end{equation}
where $\{\alpha,a,\beta,b\}$ is the set of ppE parameters with $\alpha,\beta \in \mathbb{R}$ and $a,b$ being integer multiples of $-1/3$. Sampson et al. 
\cite{sampsoncornishyunes2013} have shown that leading order corrections are sufficient to discriminate between GR and any alternative to it, while higher
order corrections play only a subdominant role. 

The existing ppE scheme has been developed in the frequency domain for a number of reasons: GW astronomy is mainly conducted using the concept of 
optimal Wiener or `matched' filtering.  Here, one assumes a theoretical waveform model based on a number of physical parameters, and correlates this 
template with a data set to test the viability of the choice of parameters.  Matched filtering works extremely well in the case where we are tasked with extracting a coherent signal buried in noise
(which is almost always the case in GW astronomy).  By carrying out the analysis in the Fourier domain, it is possible to `lift' the signal above the noise.  The important results in GW astronomy for both detection and parameter estimation require the evaluation of
noise-weighted inner products of the form
\begin{equation}
  \label{eq:innerproduct}
  \left<g|h\right> = 2 \, \int_0^\infty \frac{\tilde{g}^*(f) \tilde{h}(f) + \tilde{g}(f) \tilde{h}^*(f)}{S_n(f)} \, df,
\end{equation}
where $\tilde{g}(f)$ and $\tilde{h}(f)$ are the Fourier transforms of the time domain waveforms $g(t)$ and $h(t)$, and $S_n(f)$ is the noise spectral density of the 
detector (which we will define at a later stage).    If $S_n(f)$ is constant
across the frequency band of the detector, we could use Parseval's theorem to evaluate these inner products in the time domain.  As it is not constant for the sources we consider in this work, there is 
no closed form solution to the above integrals, and they must then be evaluated numerically in the Fourier domain.  When GW algorithms were first developed,  one had the option of generating a time domain
waveform and then carry out a numerical fast Fourier transform (FFT).  However, due to the efficiency to the algorithm and the available computers, the FFT
accounted for a large fraction of the total waveform generation time.  In this case it was clearly more advantageous to generate the waveforms directly in
the Fourier domain.  This lead to the widespread use of  the SPA in
GW parameter estimation due its low computational cost.   In most cases it has been shown to perform reasonably well for ground-based detectors such as LIGO,
although it should be pointed out that some modifications are necessary in the high mass regime \cite{dis2000,panbuonanno2008,boyle2009}. 

Nowadays, the FFT accounts for a small quantity of the total waveform generation time (normally on the order of between 3-8\%).  Therefore, as the modelling of
the binary system is conducted in the time domain, and as the GW waveforms are initially derived in the time domain, there seems to be little point in expending
the extra theoretical energy to derive either higher order approximation or alternative theory SPA waveforms.  Furthermore, several studies have shown issues with the simple (unextended) SPA waveform, especially at the high mass end and close to the last stable
orbit \cite{dis2000,panbuonanno2008,boyle2009}.  For these reasons, we chose to work in a framework of perturbed GR time domain waveforms. In order to establish a common base with
the frequency domain ppE scheme, we compute an approximate relation between both schemes. 

As matched filtering is highly sensitive to phase corrections, in this work we neglect amplitude corrections to the ppE waveforms and set $\alpha=a = 0$ in Eqn~(\ref{eq:ppE}).
In terms of the phase correction parameters $\{b,\beta\}$ as used in \eqref{eq:ppE}, we work 
with the general form corrected orbital phase
\begin{equation}
 \Phi^{(\pm)}_\text{NGR}(\Theta; b, \beta) = \Phi_\text{GR}(\Theta) \pm \Phi_c(\Theta;b, \beta),
\end{equation}
where NGR stands for `non-GR', $\Theta$ represents the dimensionless time (to be defined) and $\Phi_c(\Theta;b, \beta)$ is a corrective term to the GR phase that will also be defined at a later stage. As we stated earlier, our goal is to work in the most general context possible, allowing for all possible deviations to GR.  Therefore, in constructing our theoretical framework, we always assume that the coupling constant $\beta$ is manifestly positive, allowing us to distinguish between
positive and negative corrections to the GR waveform.

This paper is structured as follows: In Section \ref{Sec:Model} we review time and frequency domain (SPA) waveform models as they are in GR. Then, in Section \ref{Sec:Modifications},
we introduce modifications to both waveform models and set them into approximate relation. In 
Section \ref{Sec:Detection} we introduce the methodology to detect non-GR signals and carry out a parameter estimation for
the different systems.  Finally the results are presented in Section \ref{Sec:Results}.

\section{\label{Sec:Model}Waveform Models}

\subsection{Time Domain Waveform}

If we consider the quasi-circular inspiral of two non-spinning supermassive black holes with masses $m_1$ and $m_2$,
with respect to a fixed detector frame, we can define the binary unit angular momentum vector $\bm{\hat{L}}$
and the unit vector pointing from the detector to the source $\bm{\hat{n}}$. The position of the source in the
sky can then be indicated with spherical angles ($\theta,\phi$). The orientation of the binary relative to the detector
can be described by the inclination $\iota = \arccos\left[\bm{\hat{L}} \cdot \bm{\hat{n}}\right]$ and polarisation angle
$\psi = \arctan\left[\left(\bm{\hat{L}} \cdot \bm{\hat{z}} - (\bm{\hat{L}} \cdot \bm{\hat{n}}) (\bm{\hat{z}} \cdot \bm{\hat{n}})\right)/\left(\bm{\hat{n}} \cdot (\bm{\hat{L}} \times \bm{\hat{z}}) \right)\right]$.

The gravitational wave strain of the eLISA detector is, in the low frequency approximation~\cite{Cutler1997ta}, a linear
combination of $h_{+,\times}$ polarisations, weighed with the antenna patterns $F^{+,\times}$:
\begin{equation}
 \label{tdwaveform}
 h(t) = h_+[\xi(t)] F^+(t) + h_\times[\xi(t)] F^\times(t).
\end{equation}
Because of the detector motion relative to the source, a Doppler shift is introduced via a phase shifted
time parameter $\xi(t) = t - R_\oplus \sin\theta \cos(\alpha(t) - \phi)$, where $R_\oplus = 1 \text{AU}$
is the Earth-Sun distance and $\alpha(t) = 2\pi f_m t + \alpha_0$ with LISA modulation frequency $f_m = 1/\text{yr}$ 
is the orbital phase of the detector.

The antenna patterns depend on position and orientation of the source in the sky and are given for an eLISA-like detector by \cite{cornishrubbo2003, rubbocornish2004}
\begin{eqnarray}
 F^+_k(t; \psi, \theta, \phi) & = & \frac{1}{2} \left[ \cos(2\psi) D^+(t; \psi, \theta, \phi, \lambda_k) \right. \nonumber \\
                              &   & \left. - \sin(2\psi) D^\times(t; \psi, \theta, \phi, \lambda_k) \right], \\
 F^\times_k(t; \psi, \theta, \phi) & = & \frac{1}{2} \left[ \sin(2\psi) D^+(t; \psi, \theta, \phi, \lambda_k) \right. \nonumber\\
                              &   & \left. + \cos(2\psi) D^\times(t; \psi, \theta, \phi, \lambda_k) \right], \\
 \nonumber
 \end{eqnarray}
with $\lambda_1 = 0$ and $\lambda_2 = \pi/4$.
Expressions for the detector patterns 
$D^{+,\times}(t; \psi, \theta, \phi, \lambda_k)$ will not be printed here, but can be found in \cite{cornishrubbo2003}.

Many previous studies on detection and parameter estimation have used ``restricted" post-Newtonian waveforms, i.e.\ the amplitude of the 
waveform is kept at the dominant order, while the phase of the waveform is expanded to higher post-Newtonian orders.  However, a large
body of work has demonstrated that the inclusion of higher harmonic corrections to the waveform are extremely important in both the breaking
of parameter correlations and the improvement of parameter estimation~\cite{Porter2008kn,Trias2007fp,Arun2007hu,Moore1999zw,Sintes1999cg,arunetal2009}.  In fact, in some cases the estimation of luminosity distance, sky resolution
and mass determination have been shown to be improved by at least an order of magnitude due to the inclusion of the harmonic corrections~\cite{Porter2008kn}.  With
this in mind, we will use higher harmonic corrected gravitational wave polarisations up to second PN order\cite{Blanchet1996pi,blanchet2014}, i.e.
\begin{eqnarray}
 h_{+,\times} & = & \frac{2 G M \eta}{c^2 D_L} x\left[ H^{(0)}_{+,\times} + x^{1/2} H^{(1/2)}_{+,\times} \right. \nonumber \\
              &   & \label{eq:hpluscross} \left. + x H^{(1)}_{+,\times} + x^{3/2} H^{(3/2)}_{+,\times} + x^2 H^{(2)}_{+,\times} \right], \\
              \nonumber
\end{eqnarray}
where the post-Newtonian parameter $x = (GM\omega/c^3)^{2/3}$ is a function of the orbital frequency $\omega$.
The luminosity distance $D_L$ is given as a function of redshift $z$ in terms of the $\Lambda$CDM model by
\begin{equation}
 D_L = (1+z) \frac{c}{H_0} \int_0^z  \frac{dz'}{\sqrt{\Omega_R (1+z')^4 + \Omega_M (1+z')^3 + \Omega_\Lambda}},
\end{equation}
using the concurrent Planck values of $\Omega_R = 4.9 \times 10^{-5}$, $\Omega_M = 0.3086$, $\Omega_\Lambda = 0.6914$ and the Hubble constant $H_0 = 67.77$ km s$^{-1}$ Mpc$^{-1}$~\cite{Planck2013nga}. 

The waveform evolution is governed by the orbital phase and frequency of the binary, which can be expressed to 2PN order using the dimensionless time variable $\Theta(t) = \frac{\eta c^3}{5 G M} (t_c-t)$ as \cite{blanchet2014, buonannoiyer2009}
\begin{eqnarray}
  \label{eq:OmegaGR} \omega(\Theta) & = & \frac{c^3}{8 G M} \left[ \Theta^{-3/8} + \left(  \frac{743}{2688} + \frac{11}{32} \eta \right) \Theta^{-5/8} - \frac{3\pi}{10} \Theta^{-6/8} \right. \nonumber \\
   & & \left. +  \left(\frac{1855099}{14450688} + \frac{56975}{258048} \eta + \frac{371}{2048} \eta^2 \right) \Theta^{-7/8} \right], \\
  \label{eq:PhiGR} \Phi(\Theta) &=& \Phi_C - \frac{1}{\eta} \left[ \Theta^{5/8} + \left(  \frac{3715}{8064} + \frac{55}{96} \eta \right) \Theta^{3/8} - \frac{3\pi}{4} \Theta^{2/8} \right. \nonumber \\
  & & \left. + \left( \frac{9275495}{14450688} + \frac{284875}{258048} \eta + \frac{1855}{2048} \eta^2 \right) \Theta^{1/8} \right]. \\
  \nonumber
\end{eqnarray}
Multiples of the orbital phase are then manifest in the harmonic coefficients $H^{(n)}_{+,\times}(\Phi,\iota,m_1,m_2)$.

\subsection{Stationary Phase Approximation}

To work directly in the Fourier domain, one requires an analytic form of the Fourier transform $\tilde{h}(f) = \int e^{2\pi i f t} h(t) \, dt$. Many previous studies in the field of GW data analysis have employed the stationary phase approximation 
which gives a reasonable approximation of the Fourier integral, respecting some limitations toward high masses and close to the last
stable orbit.  In constructing the SPA, one assumes that because of the rapid oscillatory nature of the integrand, the Fourier integral averages to
zero, except at points where the phase function has an extremum.  By expanding the phase in a Taylor series around the stationary point, the integral
takes on the form of a Fresnel integral with a standard solution.

Only considering the dominant harmonic of Eqn~(\ref{tdwaveform})
with $H^{(0)}_+ = -(1+\cos^2\iota) \cos2\Phi$ and $H^{(0)}_\times = -2\cos\iota \sin2\Phi$,
one ends up with (see e.g. \cite{yunesetal2009,langhughes2006})
\begin{eqnarray}
 \label{SPAWaveform}
 \tilde{h}(f) & = & \sqrt{\frac{5}{96}} \pi^{-2/3} \frac{c}{D_L} \left(\frac{G\mM}{c^3}\right)^{5/6} A_{\text{pol}}[t(f)] \, f^{-7/6} \nonumber \\
 & & \times e^{i(\Psi(f) - \varphi_{\text{pol}}[t(f)] - \varphi_D[t(f)])},
\end{eqnarray}
with the SPA phase given by
\begin{equation}
 \label{PsiSPA}
 \Psi(f) = 2\pi f t(f) - 2\Phi[t(f)] - \frac{\pi}{4},
\end{equation}
and polarisation amplitude and phase
\begin{eqnarray}
 A_{\text{pol}} & = & \sqrt{(1 + \cos^2\iota)^2 \, F^+(t)^2 + 4 \cos^2\iota \, F^\times(t)^2}, \\
 \varphi_{\text{pol}} & = & \text{atan2} \left[ 2 \cos\iota \, F^\times(t), (1 + \cos^2\iota)^2 \, F^+(t) \right]. \\
 \nonumber
 \end{eqnarray}
The Doppler phase correction caused by detector motion is given by 
\begin{equation}
 \varphi_D = 2\pi f \frac{R_{\oplus}}{c} \sin\theta \cos(\alpha[t(f)]-\phi)].
\end{equation}
Finally, the time evolution $t(f)$ is given by the TaylorT2 timing function which up to 2PN order is defined by \cite{buonannoiyer2009}
\begin{eqnarray}
 t(f) & = t_c & - \frac{5 G M}{256 \eta c^3 x^4} \left[ 1 + \left( \frac{743}{252} + \frac{11}{3} \nu \right) x - \frac{32}{5} \pi x^{3/2} \right. \nonumber \\
 & & \left.  + \left( \frac{3058673}{508032} + \frac{5429}{504} \eta + \frac{617}{72} \eta^2\right) x^2 \right].
\end{eqnarray}
The SPA has been shown to be reasonably accurate as long as the binary inspiral is in the adiabatic regime, i.e.\ as long as $\frac{d(\log{a(t)})}{dt} \ll \frac{d\Phi}{dt}$ 
and $\frac{d^2\Phi}{dt^2} \ll \left( \frac{d\Phi}{dt}\right)^2$, where $a(t)$ is the amplitude of the GW \cite{droz1999, damournagartrias2011}.

\section{\label{Sec:Modifications}A Relation Between Waveform Models Modified in Time and Frequency Domains }

\subsection{\label{Sec:WaveformRelations}A Relation between Modified Waveforms in Time and Frequency Domains}

To derive the perturbed time domain waveform, we start in the same manner as \cite{yunespretorius2009} and introduce a leading order correction to the time domain orbital phase, albeit
based a little bit more on the general form of the GR orbital phase given by Eqn~\eqref{eq:PhiGR}:
\begin{equation}
 \label{eq:PhiModAlpha}
 \Phi_\text{NGR}(\Theta) = \Phi_\text{GR}(\Theta) \pm \frac{1}{\eta} \kappa_i(b,\beta) \, \Theta^{\frac{5-2i}{8}},
\end{equation}
where $\kappa(b,\beta) \in \mathbb{R}$ and $i \in \{0,1/2,1,3/2,2\}$ allow the corrections to enter somewhere between 0PN and 2PN. We choose not to consider corrections above 2PN
and below 0PN in this work: we disregard `negative' PN terms such as a `-1PN' dipole moment correction for simplicity and  
because for the main class of theories exhibiting dipole radiation (scalar-tensor theories), SMBHB inspirals
are not expected to emit dipole radiation \cite{lang2014}. Furthermore, corrections coming in below 1PN order are better
constrained using solar system tests and binary pulsar observations \cite{yuneshughes2010, cornishsampson2011}.

To compare our results with the considerable effort that has already been done in the field, we relate Eqn~\eqref{eq:PhiModAlpha} to a leading order phase-only correction ppE scheme.
Eqn~\eqref{eq:ppE} implies the phase correction
\begin{equation}
 \label{eq:PsiPPE}
 \Psi_\text{NGR}(b,\beta; \, u) = \Psi_\text{GR}(u) \mp \beta \, u^b.
\end{equation}
To relate Eqns~\eqref{eq:PhiModAlpha} and \eqref{eq:PsiPPE}, we construct a time domain waveform, that, after a numerical Fourier transform, approximately
reproduces a frequency domain waveform with a modified phase as in Eqn~\eqref{eq:PsiPPE}. To this end, we introduce sub-leading terms to Eqn~\eqref{eq:PhiModAlpha} which then becomes
\begin{equation}
\label{eq:modelPhi}
 \Phi_{\text{NGR}}(\Theta) =  \Phi_{\text{GR}}(\Theta) \pm \frac{1}{\eta} \sum_i \kappa_i \Theta^{\frac{5-2i}{8}}.
\end{equation}
The next question to answer is what set of $\{\kappa_i\}$ leads to a time domain waveform that is consistent with
a modified SPA waveform with phase \eqref{eq:PsiPPE} .
The necessary relations can be found by reconsidering the steps that led us to the SPA waveform.
Eqn~\eqref{PsiSPA} can, in terms of the reduced frequency $u$, be written as
\begin{equation}
 \Psi(u) = 2 \left[ \frac{c^3}{G \mM} t(u) \, u - \Phi(u) \right] - \frac{\pi}{4}.
\end{equation}
Since the frequency derivative of the orbital phase can be expressed as
\begin{equation}
 \frac{d\Phi}{du}(u) = \frac{dt}{du}(u) \, \frac{d\Phi}{dt}[t(u)] = \frac{dt}{du} \frac{c^3}{G \mM} u,
\end{equation}
the frequency derivative of the SPA phase reduces then to the simple expression
\begin{equation}
 \frac{d\Psi}{du} = 2 \frac{c^3}{G \mM}\,  t(u).
\end{equation}
This enables us to write the time-of-frequency function in the simple form
\begin{equation}
\label{tu}
 t(u) = \frac{1}{2} \frac{G\mM}{c^3} \frac{d\Psi}{du}.
\end{equation}
Similarly, we can write the inverse relation as
\begin{equation}
 \label{ut}
 u(t) = \frac{G\mM}{c^3} \frac{d\Phi}{dt}.
\end{equation}
One should remember that for Eqn~\eqref{tu} we have only considered the dominant harmonic, while Eqn~ \eqref{ut} uses no such assumption.
The functions $t(u)$ and $u(t)$ inherit the corrections applied to $\Psi$ and $\Phi$ in the previous section in a simple manner.
Since we require $u[t(u)] = u$, relations between the time domain and the SPA phase coefficients can be 
computed at ease.
In terms of $\Theta$ the above expressions can be written as
\begin{equation}
\label{uTheta}
 u(\Theta) = \frac{G\mM}{c^3}  \frac{d\Theta}{dt} \frac{d\Phi}{d\Theta}, \qquad \Theta(u)  = \Theta[t(u)]. \nonumber \\
\end{equation}
The coefficients of $\Psi(u)$, given the coefficients for $\Phi(\Theta)$, can thus be computed by evaluating the equation
\begin{equation}
 \label{eq:utu}
 u[ \Theta( u ) ]_{\text{2PN}} = u \, \left(1+\sum_{k=0}^4 u^{k/3} \mathcal{A}_k \right) = u,
\end{equation}
expanded up to 2PN order in $u$. For the non-GR time domain and SPA phases, we do this at linear order in $\kappa_i$ and $\beta$,
assuming that the corrections are small enough.
Setting the $\mathcal{A}_k$ to zero, the resulting system can then be solved for $\kappa_i(b, \beta)$. 

We restrict $b$ and $i$ such that corrections can come in only between 0PN and 2PN order, in detail $b \in \{-5/3, -4/3, -1, -2/3, -1/3\}$
and $i \in \{0,1/2,1,3/2,2\}$; this corresponds to 0PN, 0.5PN, 1PN, 1.5PN, 2PN corrections, respectively.

While we fix one particular value of $b$, we always allow the full sum of corrections proportional to $\kappa_i$ in the time domain phase.
This enables us to compute $\kappa_i(b,\beta)$ for each value of $b$; the results are listed in Table \ref{Table:alphabeta}.
\squeezetable
\begin{table}
\caption{\label{Table:alphabeta} Impact of different frequency domain corrections to the phase on the parameter space of time domain corrections $\kappa_i(b,\beta)$.}
\begin{ruledtabular}
\begin{tabular}{l||l|l|l|l|l}
b	  & -5/3 & -4/3 & -1 & -2/3 & -1/3 \\
\hline\hline
$\kappa_{0}$  			& $16 \beta$ & 0 & 0 & 0 & 0 \\
\hline
$\kappa_{1/2}$  		& 0 & $8 \beta \eta^{1/5}$ & 0 & 0 & 0  \\
\hline
$\kappa_{1}$  			& $-16 \beta \Phi_1$ & 0 & $4 \beta \eta^{2/5}$ & 0 & 0 \\
\hline
$\kappa_{3/2}$  		& $-\frac{32}{3} \beta \Phi_{3/2}$ &  $-\frac{32}{5} \beta \Phi_1 \eta^{1/5}$ & 0 & $2\beta \eta^{3/5}$ & 0 \\
\hline
$\kappa_{2}$ 	 		& $16 \beta \left(\frac{4}{5}\Phi_1^2 - \frac{1}{3}\Phi_2 \right)$ & $-\frac{64}{15} \beta \Phi_{3/2} \eta^{1/5}$ & $-\frac{12}{5} \beta \Phi_1 \eta^{2/5}$ & 0 & $\beta \eta^{4/5}$ \\
\end{tabular}
\end{ruledtabular}
\end{table}

As expected, the $\kappa_i$ are always proportional to $\beta$.
If the frequency domain phase correction enters at $n$PN order, $n=(3b+5)/2$, then $\kappa_{i<n} = 0$ and the remaining $\kappa_i$ 
are proportional to $\eta^{\frac{2n}{5}}$. Also one can see that the lowest order correction $\kappa_i$ then has the numeric prefactor 
$2^{4-2i}$.
Moreover, in each diagonal term, the coefficients are proportional to $\Phi_0=1$, in the first off-diagonal to $\Phi_{1/2} = 0$, 
in the second off-diagonal to $\Phi_1$ and in the third off-diagonal to $\Phi_{3/2}$. In the  fourth off-diagonal element, terms proportional
both to $\Phi_1^2$ and $\Phi_2$ can appear. 

It is important to stress that these are approximate relations, i.e.\ we do not expect the time domain waveform to match the SPA waveform
perfectly. This is because we expect time domain and SPA waveforms to differ near the last stable orbit due to approximation errors in the SPA.
 However, the waveform match is more than good enough to accomplish our task of constraining the `coupling constant' $\beta$.

By the same argument as in \cite{sampsonyunescornish2013}, we assume that leading order corrections to the time domain phase are already enough
to discriminate between GR and competing theories. This can be justified through the fact that the next-to-leading order term is always of order
1PN away from the leading order term (see Table \ref{Table:alphabeta}),
hence we usually end up with $\left[\kappa_{i+1} \Theta^{\frac{5-2(i+1)}{8}}\right] / \left[\kappa_i \Theta^{\frac{5-2i}{8}}\right] \sim \Theta^{-\frac{1}{4}}$.
At an orbital separation of $R_\text{max} = 7 \, GM/c^2$, $\Theta^{-1/4}$ usually takes a numerical value of around $0.7$, while away from the last stable
orbit it will be much smaller.

Therefore, to leading order, taking only the diagonal terms in Table \ref{Table:alphabeta} into account, we can write our non-GR corrected phase as
\begin{equation}
 \Phi_{\text{NGR}}^{(\pm)}(b,\beta;\,\Theta) = \Phi_\text{GR}(\Theta) \pm  2^{-1-3b} \beta \, \eta^{3b/5} \Theta^{-3b/8},
\end{equation}
where $i = \frac{3b+5}{2}$ has been used.   This expression corresponds to the modified SPA phase
\begin{equation}
\label{eq:PsiPPEpm}
 \Psi_{\text{NGR}}^{(\pm)}(b,\beta;\,u) = \Psi_{\text{GR}}(u) \mp \beta u^b.
\end{equation}
We should note here that the sign of the corrections in the time domain are reversed from those in the frequency domain.

In our study, we assume that $\beta$ is manifestly positive for two reasons.  The first is that we would like to
be able to differentiate between positive and negative corrections to the GR phase.  The second is a numerical reason : when carrying out a parameter
estimation study, we will need to calculate the Fisher information matrix (FIM).  As we expect $\beta$ to be quite small for certain theories, the FIM is
numerically more stable if we chose $\ln(\beta)$ as the working parameter rather than $\beta$ itself, thus requiring a positive value.

Given the non-GR corrected phase, we can now also write down the modified orbital frequency evolution
\begin{equation}
 \omega_{\text{NGR}}^{(\pm)}(b,\beta;\Theta) = \omega_\text{GR}(\Theta) \pm  2^{-4-3b} \frac{3\beta}{5} \frac{c^3}{GM}\, \eta^{3b/5+1} \Theta^{-3b/8-1}.
\end{equation}
Using these expressions,  our next goal is to compute for what size of $\beta$ the eLISA detector is able to distinguish between GR and a competing theory.
This is done in Section \ref{Sec:Results} for a total of ten different cases: five values of $b$ (as defined above) for $\Phi_\text{NGR}^{(+)}$ and $\Phi_\text{NGR}^{(-)}$ individually.   In the next subsection, we discuss the limits that we choose to set on $\beta$, which will further be used as priors in our study.

\subsection{Suggested Limits for the Coupling Constant $\beta$}

The possible values of $\beta$, and hence $\kappa_i(b, \beta)$, have to be limited for two reasons:
firstly, due to the fact that we make the assumption that we are working with a perturbed GR waveform. By doing so, one can then, after successful detection, conduct 
a further analysis of the recovered signal and check it against different theories. If the corrections to the GR waveforms are too large (which
they could well be after 1PN order corrections where they there they are essentially unconstrained), then GR waveforms will fail to detect the signal.  Secondly, in the
upcoming chapters, we use a Markov Chain Monte Carlo algorithm (MCMC) to not only confirm detection, but also to conduct a Bayesian inference.  For the MCMC to work,  we need to place priors on $\beta$, as a function of the theory described by different values of $b$.

\begin{figure}[t!]
 \includegraphics[width=\columnwidth]{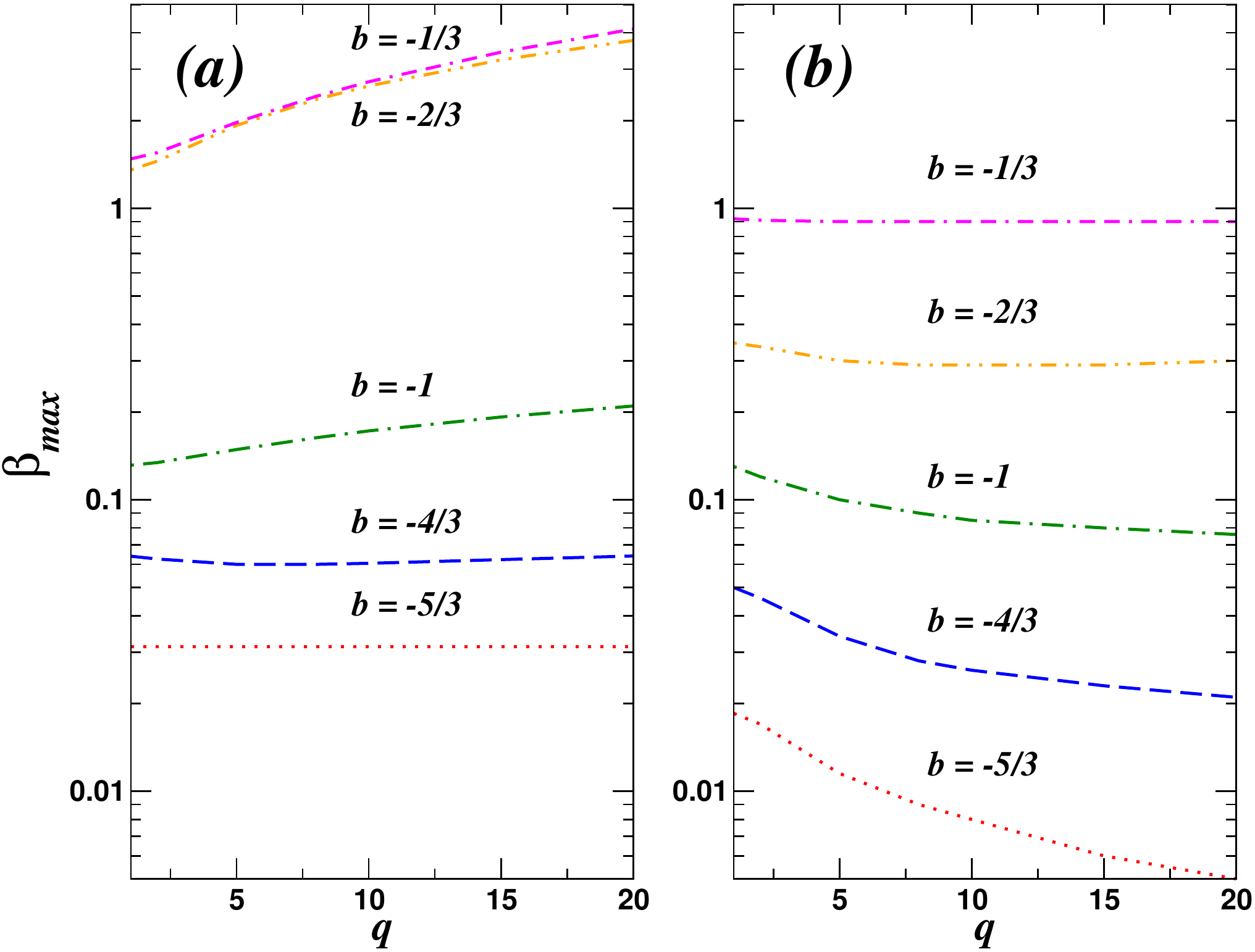}
 \caption{\label{Fig:upperlimits} Maximum allowed values for $\beta$ as a function of mass ratio $q$ given the constraint of $\kappa_i/\Phi_i \leq 0.5$ for different
 alternative theories.  Cell (a) represents non-GR corrections with positive sign, while cell (b) represents non-GR corrections with negative sign, with
 $b=-1/3$ (red dotted line), $b=-2/3$ (blue dashed line), $b=-1$ (green dot-dashed line), $b=-4/3$ (orange double dot-dashed line) and $b=-5/3$ (magenta double dash-dot line).}
\end{figure}

In order to limit the perturbation of the GR waveform, one should demand that $\kappa_i$ is small with respect to the fiducial orbital phase
coefficients (let us call them $\Phi_i$) in  Eqn~\eqref{eq:PhiGR}. One issue is that not every correction has its own fiducial GR coefficient. Since there is no 0.5PN
term in the GR phase evolution, we have no way of constraining the corresponding $\kappa_{1/2}$ coefficient. In this case, we simply make sure that all the other non-zero subleading coefficients in Eqn~\eqref{eq:modelPhi} (i.e.\ $\kappa_{3/2}$ and $\kappa_{2}$) stay within a certain limit. For this study, we choose the limit to be 50\% of the GR coefficient value:
\begin{equation}
 \underset{i}{\text{max}} \, \left| \frac{\kappa_i}{\Phi_i} \right| < 0.5.
\end{equation}
This results in a limit on $\beta$ as a function of mass ratio $q$, depending on what order of approximation a correction enters due to a theory with a particular value of $b$. 
We found that for all cases, regardless of whether the corrections have a positive or negative sign,  except for $b=-4/3$, the first non-zero $\frac{\kappa_i}{\Phi_i}$ dominates the others.
Interestingly, for  $b=-4/3$, $\frac{\kappa_2}{\Phi_2}$ dominates $\frac{\kappa_{3/2}}{\Phi_{3/2}}$.
The resulting upper limits on $\beta$ are plotted as a function of the mass ratio in Fig~\ref{Fig:upperlimits}.

In cell (a) we plot the limits on $\beta$ for theories with positive sign corrections, while in cell (b) we have the maximum values of $\beta$ for theories with
negative corrections.  In Table~\ref{Table:upperlimits} we also provide an analytic form for the limit of $|\beta|$ for the different values of $b$.   As these 
values are correct for the magnitude of the coupling constant, it would suggest that the limits plotted in Fig~\ref{Fig:upperlimits} should be identical regardless
of the sign of the corrections.  However, we can clearly see that this is not the case.  The reason for the discrepancy between the plot and the table is due to 
pathologies in the evolution of the non-GR orbital frequency.  For all cases, our goal is to evolve the binary systems to a minimum separation of $R=7M$.  This 
works in all cases when we consider positive corrections to the phase.  However, when we introduce a negative sign correction, the gradient of the non-GR
orbital frequency changes sign before reaching $7M$.  As a consequence, we are then required to terminate the waveform evolution at the point where
$d\omega_{NGR} /dt = 0$.  Thus, the maximum limits for $\beta(q)$ plotted in cell (b) correspond to waveforms terminated at a separation of $R(d\omega_{NGR} /dt = 0)$.  We should point out that as we go to smaller values of $\beta$ in each theory, we do recover a situation where the waveforms are once more terminated
at $R=7M$.

\begin{table}[t]
\caption{\label{Table:upperlimits} Upper limits on $\beta$ for different powers of $b$, assuming that the $\kappa_i(b,\beta)$ are at maximum 50\% of the fiducial GR orbital phase parameters $\Phi_i$.}
\begin{ruledtabular}
\begin{tabular}{l|ll}
 $b = -5/3$ (0PN) & $|\beta| < 0.03125$ & \\
 \hline
 $b = -4/3$ (0.5PN) & $|\beta| < \frac{15}{128 \eta^{1/5}} \left|\frac{\Phi_2(\eta)}{\Phi_{3/2}}\right|$ & \\
 \hline
 $b = -3/3$ (1PN)  & $|\beta| < \frac{|\Phi_1(\eta)|}{8 \eta^{2/5}}$ & \\
 \hline
 $b = -2/3$ (1.5PN) & $|\beta| < \frac{|\Phi_{3/2}|}{4 \eta ^{3/5}}$ & \\
 \hline
 $b = -1/3$ (2PN) & $|\beta| < \frac{|\Phi_{2}(\eta)|}{2 \eta ^{4/5}}$ & \\
\end{tabular}
\end{ruledtabular}
\end{table}

\section{\label{Sec:Detection}Detecting non-GR signals with GR Waveforms}

\subsection{Detector Configuration}
\begin{figure}[t!]
 \includegraphics[width=\columnwidth]{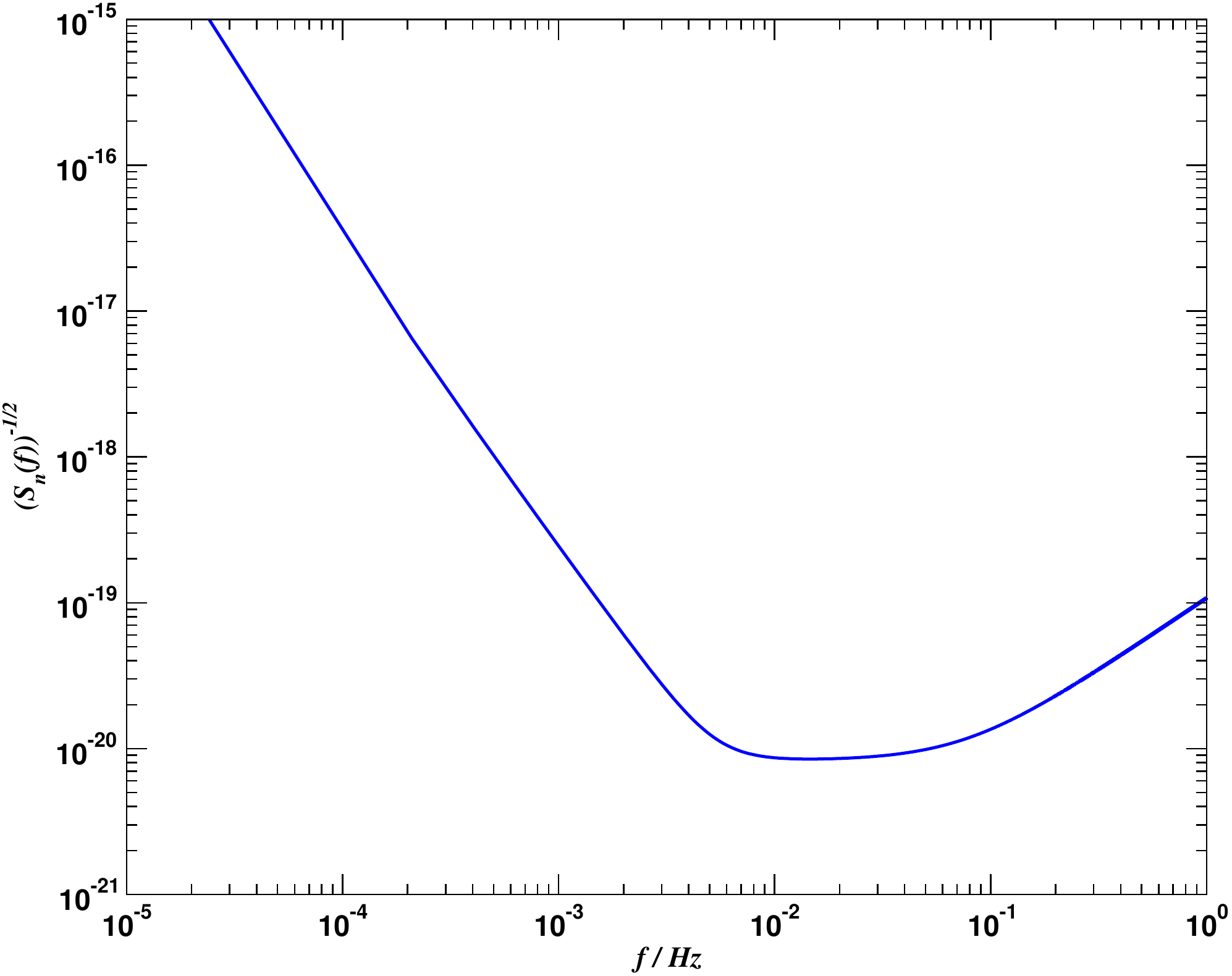}
 \caption{\label{fig:noise}Instrumental noise model for a $10^9$m arm eLISA configuration.}
\end{figure}

For this study we assume an eLISA configuration, where the space-craft are separated by $10^6$ kms
and operate using four laser links.  In this configuration, the observatory can be interpreted as a single channel
Michelson interferometer.  This corresponds to the eLISA configuration accepted as a candidate for the
ESA Cosmic Vision L3 mission concept \cite{eLISAWhitepaper,ngoscience}.  The noise power spectral density for the eLISA observatory
can be modelled using the form
\begin{eqnarray}
 S_n^\text{instr}(f) & = & \frac{1}{4L^2} \left[ S_n^{fxd} + 2 S_n^{\text{pos}} \left( 2+\cos^2\left(\frac{f}{f_*}\right)\right)  \right. \nonumber \\
 & &  + 8 S_n^{\text{acc}} \left( 1+ \cos^2\left(\frac{f}{f_*}\right)\right)  \nonumber \\
 & & \left. \times \left( \frac{1}{(2\pi f)^4} + \frac{(2\pi 10^{-4})^2}{(2\pi f)^6}\right) \right]
\end{eqnarray}
where $L=10^9$m is the arm-length of the particular LISA configuration, $S_n^{\text{pos}} (f)= 1.21\times10^{-22} \text{m}^2/\text{Hz}$ is the position noise,
$S_n^{\text{acc}} (f)= 9\times10^{-30}\text{m}^2/(\text{s}^4\text{Hz})$ is the acceleration noise, $S_n^{fxd} = 6.28\times10^{-23}\text{m}^2/\text{Hz}$ is a frequency independent
 fixed level noise in the detector
 and $f_* = 1/(2\pi L)$ is the mean
transfer frequency.  We plot this noise curve in Fig.~\ref{fig:noise}.

\subsection{Bayesian Inference and MCMC}
Our goal in this work is twofold : we are first of all interested in testing the capability of GR templates in detecting non-GR signals for differing
values of $(b,\beta)$.  Once we have confirmed the regions of ``detectability", the next question to answer is to what values of $\beta$ are we capable of
resolving the system parameters without having to resort to using non-GR templates for different values of $b$.

In each case, we inject a non-GR signal into random Gaussian instrumental noise.  Given a detector response $s(t) = h_{NGR}(t;\bm{\lambda}, b, \beta) + n(t)$, where
$h_{NGR}$ is our corrected signal and $n(t)$ is random instrumental noise,  and a GR template $h(t;\bm{\lambda})$ with the 9-dimensional binary physical parameter vector 
$\bm{\lambda} = (m_1,m_2,\theta,\phi,\iota,\psi,D_L,t_c,\Phi_c)$, we define the likelihood function
\begin{equation}
 {\mathcal L}(\bm{\lambda})= C \exp\left[-\frac{1}{2} \left<s-h(\bm{\lambda})|s-h(\bm{\lambda})\right> \right],
 \label{eqn:likelihood}
\end{equation}
where $C$ is a normalisation constant.  The posterior distribution for the parameters
$\bm{\lambda}$ is then given by Bayes' theorem
\begin{equation}
 p(\bm{\lambda} | s) = \frac{\pi(\bm{\lambda}) \, {\mathcal L}(\bm{\lambda})}{p(s)},
\end{equation}
where $\pi(\bm{\lambda})$ is the prior distribution of the binary parameters,
and $p(s)$ is the marginal likelihood or model evidence.

\subsubsection{Markov Chain Monte Carlo}
As it is not our goal to conduct a full search over the entire possible parameter space for massive black hole binaries, we make the fundamental assumption that we have been able to narrow the search region by some other means.  At this point we use a non-Markovian Metropolis-Hastings MCMC to narrow the
search space even further.  When using a standard MCMC, one picks a starting point in the parameter space, and by using a tailored
proposal distribution, proposes a jump to another part of the parameter space with parameters $\bm{\lambda'}$ using a proposal distribution of choice 
$q(\,|\,)$.  One then compares the new and old points in parameter space by evaluating the Metropolis-Hastings ratio
\begin{equation}
  H = \frac{\pi(\bm{\lambda'}) \, {\mathcal L}(\bm{\lambda'}) q( \bm{\lambda} | \bm{\lambda'})}{\pi(\bm{\lambda}) \, {\mathcal L}(\bm{\lambda}) q( \bm{\lambda'} | \bm{\lambda})}.
 \end{equation}
We should mention here that in reality we work with a reduced log likelihood
\begin{equation}
\ln {\mathcal L} = \left<s|h(\bm{\lambda})\right> - \frac{1}{2}\left<h(\bm{\lambda})|h(\bm{\lambda})\right>,
\end{equation}
that is achieved by expanding the exponent of Eqn~(\ref{eqn:likelihood}) and dropping the $-1/2\left<s|s\right>$ term as it is common to both the numerator
and denominator of the Metropolis-Hastings ratio, and therefore, unimportant.

If we work with a Hessian MCMC, we can use a multivariate Gaussian based on the FIM as our proposal distribution~\cite{Cornish2006ms}.  The FIM is defined as
 \begin{equation}
 \Gamma_{\mu\nu} = \left< \frac{\partial h_{GR}}{\partial \lambda^{\mu}} \bigg| \frac{\partial h_{GR}}{\partial \lambda^{\nu}} \right>,
\end{equation}
where we explicitly specify that we are constructing the FIM using GR templates.

\begin{figure}
\begin{minipage}{\columnwidth}
\includegraphics[width=3.2in]{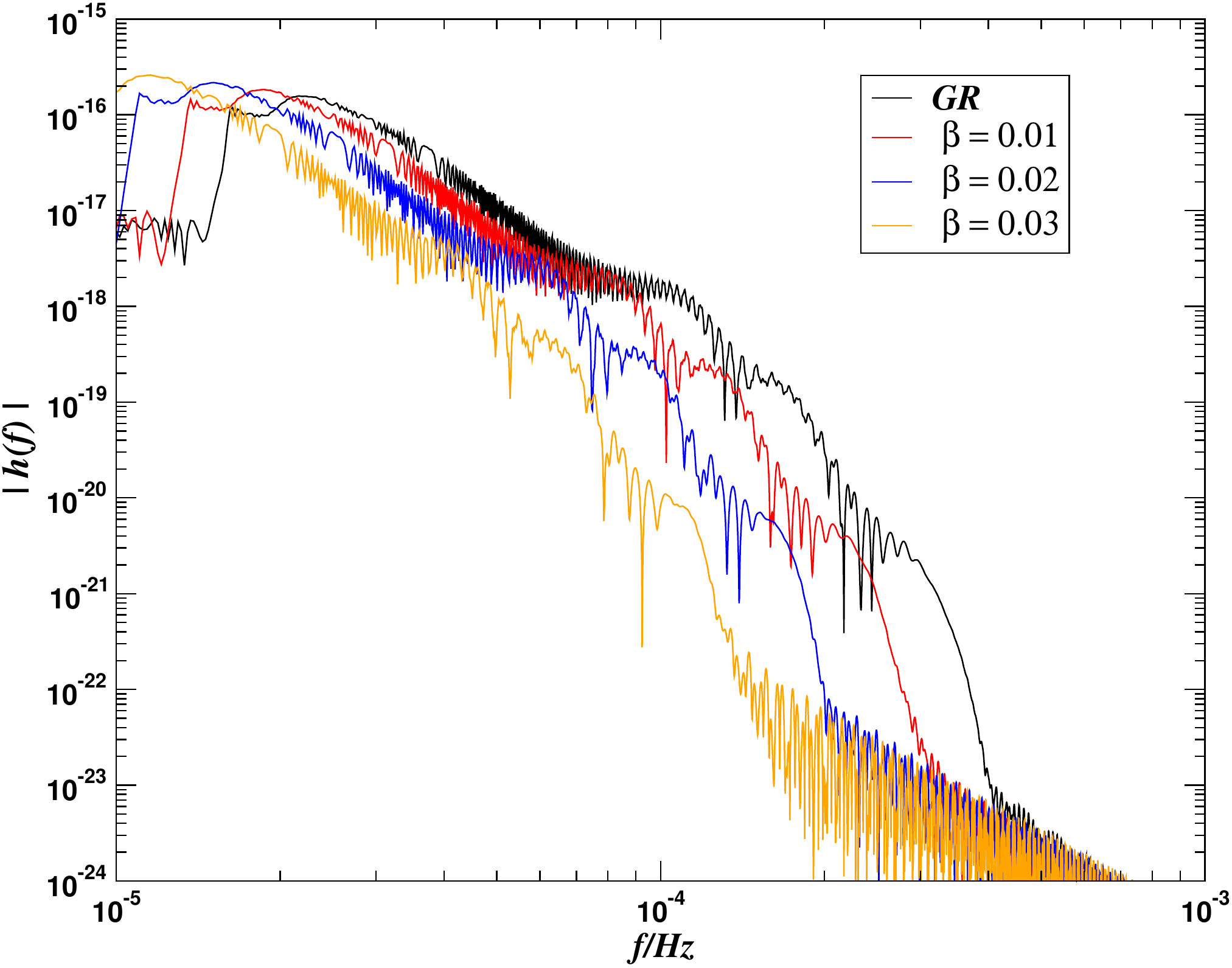}
\end{minipage}
\begin{minipage}{\columnwidth}
\includegraphics[width=3.2in]{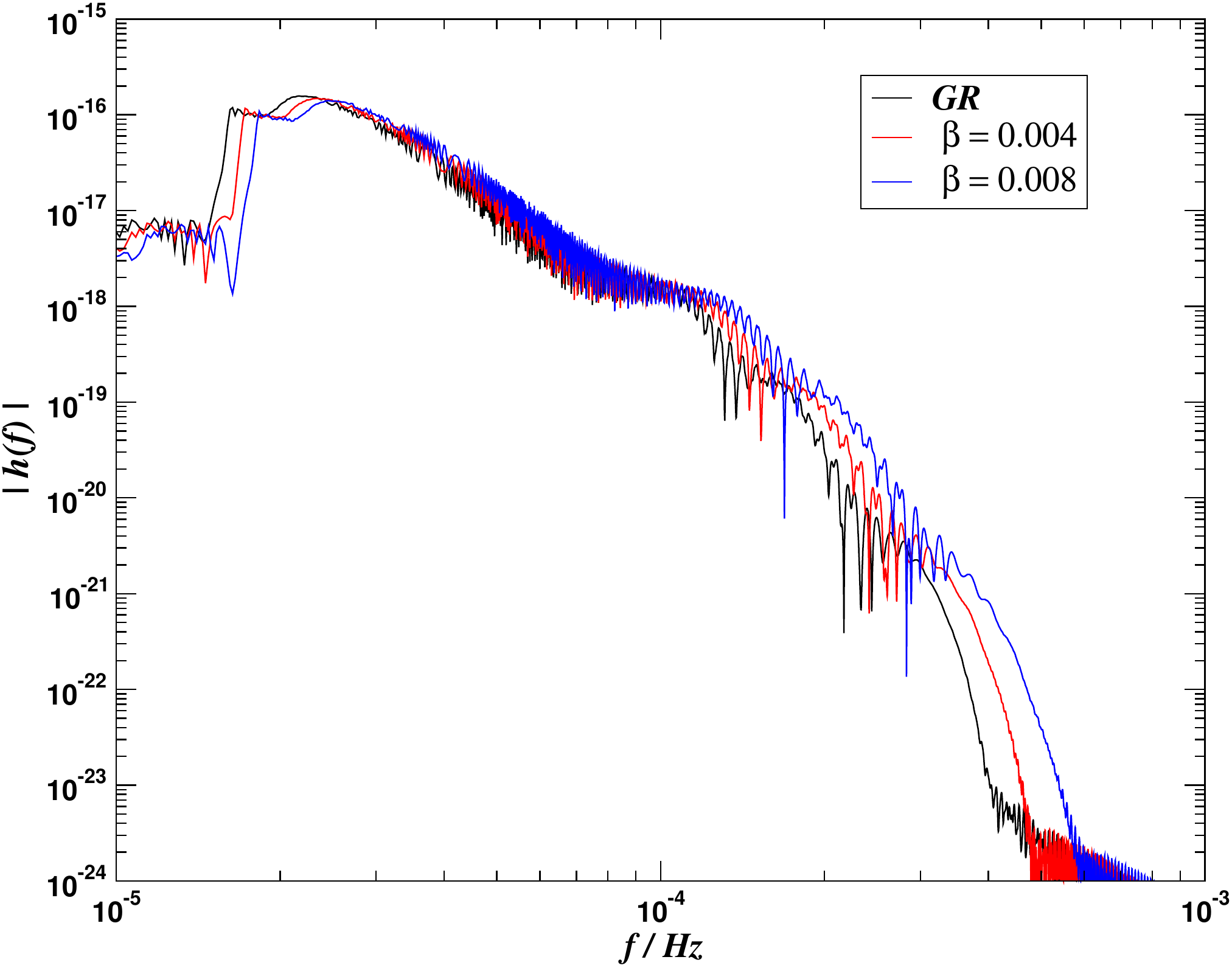}
\end{minipage}
\caption{Power spectra for both GR and non-GR waveforms assuming a SMBHB with individual source-frame masses of $10^7 - 10^6\,M_{\odot}$ at $z=1$, for
an alternative theory with $b=-5/3$, and for differing values of $\beta$.  The top figure represents positive non-GR corrections, while
the bottom figure displays negative non-GR corrections.}
\label{fig:powerspectra}
\end{figure}

As we expect there to be a difference between the GR and non-GR waveforms for some particular theories, even if we inject GR templates with the true parameter values, we do not
expect the starting points to be close to the final solution due to the mismatch between the two waveform models.  This will be especially evident in the investigation of the $b=-5/3$ and $b=-4/3$ theories.  In Fig.~\ref{fig:powerspectra} we plot the power spectra for both the GR waveforms and for the non-GR waveforms in the $b=-5/3$ theory, assuming a SMBHB with masses of $10^6-10^7\,M_{\odot}$ at $z=1$.  The top image corresponds to positive corrections, while the bottom image represents negative corrections.  In each case we plot a range of different values of
the coupling constant $\beta$, with the maximum value corresponding to the limits derived earlier.   If we first investigate the positive corrections, we can see that
the increased values of $\beta$ shift the waveforms to lower frequencies.    In the case of $\beta=0.03$,
we can see that the dominant harmonic has started to drift below the lower frequency cutoff of the detector at $f=10^{-5}$ Hz.  For a GR template to 
detect this signal, it would first have to move its total mass to a higher value, and then change its mass ratio to fit the spread of the power spectrum.
An investigation of the corresponding time domain waveforms shows that the higher the value of a positive $\beta$ correction, the faster the waveform reaches the termination radius of $R=7M$.  

In the lower image, we observe that the higher the value of $\beta$, the more the waveforms are shifted to higher frequencies as compared with the GR waveform.  Again
an investigation of the corresponding time domain waveforms demonstrates that it takes longer in the case of negative corrections for the waveforms to reach $R=7M$.  For other theories, while the patterns are the same for both positive and negative corrections, the correction at each lower value of $b$ induces a correction into the
GR phase and frequency at higher PN orders.  This implies that the deviations from a GR waveform become smaller as we go to lower values of $b$.

Due to the large possible mis-match between the GR and non-GR waveforms, it turns out that the final ``detection" parameters for many of these systems lie many hundreds of sigma away from the input parameters.  Therefore, while we are starting relatively close to the final solution in the parameter space, a short search 
phase is required for the MCMC to converge.  To achieve convergence, we use two types of annealing schemes.  The goal of annealing is to smoothen irregularities in the likelihood surface
allowing the Metropolis-Hastings chain to converge on a solution quicker.  To accomplish this, we replace the factor of 1/2 in Eqn~(\ref{eqn:likelihood}) by an inverse temperature $\gamma = 1/(2T)$, where $T$ is the temperature of the likelihood surface.  When the temperature is large, the likelihood surface is smoother and flatter.  One then slowly cools the temperature from some initial value $T_{ini}$ to $T=1$, thus returning $\gamma$ to a value
of 1/2, with the hope that the chain has now converged to the global solution.  A problem with this method is choosing firstly, the initial temperature and secondly, the
cooling rate.  It was shown that the first of these problems can be overcome by allowing the chain to control the injected heat itself.  This method, called
thermostated annealing~\cite{Cornish2006ms} injects a heat according to the rule
\begin{equation}
\gamma = \left\{ \begin{array}{ll} \frac{1}{2} & 0\leq \rho\leq \rho_0 \\ \\ \frac{1}{2}\left(\frac{\rho}{\rho_0}\right)^{-2} & \rho > \rho_0  \end{array}\right. ,
\label{eqn:thermann}
\end{equation}
where we define the signal to noise ratio (SNR) $\rho$ as
\begin{equation}
\rho = \frac{\left<s|h\right>}{\sqrt{\left<h|h\right>}}
\end{equation}
and $\rho_0$ is a SNR threshold value of choice.  As $\ln{\mathcal L} \propto \rho^2$, the second quantity in Eqn~(\ref{eqn:thermann}) is nothing more than a normalised log-likelihood.  For this study,
we choose $\rho_0 = 5$.  We run the thermostated annealing phase for the first $2\times10^4$ iterations, whereupon we use a standard simulated annealing
phase~\cite{Cornish2006dt,Cornish2006ry}
\begin{equation}
\gamma = \left\{ \begin{array}{ll} \frac{1}{2}10^{-\xi\left(1-\frac{i}{t_{cool}}\right)} & 0\leq i\leq t_{cool} \\ \\ \frac{1}{2} & i > t_{cool}  \end{array}\right.,
\end{equation}
to cool the surface.  In the above expression, $\xi = log_{10}(T_{th})$, where $T_{th}$ is the temperature at the end of the 
thermostated annealing phase, $i$ is the iteration number, and $t_{cool}$ is the cooling schedule for the simulated annealing, which we take to be
$10^4$ chain iterations.  At this point we begin a standard Hessian MCMC to estimate the recovered parameters. 

As the starting point of the chain, even though they are the true input parameters, may lie many hundreds of sigma from the final solution, the
chain may start in a deep valley in the parameter space.  In this situation, it can take a long time for the chain to
move onto a peak.  To accelerate this process, we use a maximisation over the time of arrival $t_c$ as this helps us to match the final most relativistic cycles of
the waveform.  This is done by calculating the correlation
between the data and template, and searching for the maximum of the correlation.  During the annealing phases of the algorithm, we maximise
 over $t_c$ at every iteration while the reduced log-likelihood $\ln {\mathcal L} \leq 0$.  Once the log-likelihood is positive, we then carry out a maximisation every ten 
 iterations, otherwise, we allow the chain to search freely over $t_c$. 

\subsubsection{Setting priors for the MCMC}
Finally, we impose the following priors on a subset of the physical parameters : using a flat prior, we constrain the maximum possible redshifted mass to be less than 
$1.163\times10^8\,M_{\odot}$.  This ensures that the minimum last stable orbit frequency for higher harmonic waveforms is approximately at $5\times10^{-5}$ Hz.  For the 
symmetric mass ratio, we confine our search to the flat prior between $0.01\leq\eta\leq 0.25$,  corresponding to a mass ratio domain of
$1 \leq q\leq 100$.  For the luminosity distance, we use a flat prior given by $7.7\times10^{-4}\leq D_L / Gpc \leq 110$.  The lower bound on this prior assumes that the closest a SMBHB can exist is in the
M31 (Andromeda) galaxy.  The upper bound corresponds to a redshift of $z\sim10$.  Finally, as we assume an observation period of one year, we restrict the search over
time of coalescence to $0.2\leq t_c/yrs \leq 0.99$.  All angular parameters are allowed to vary over their natural ranges.

In each run, we start the MCMC algorithm at the true input values as our goal is not only
detection of the non-GR signal, but also an estimation of parameters.

To investigate the capabilities of the eLISA detector, we chose five test sources with mass combinations of $(m_1, m_2) = \{(1.1,1)\times10^6, 
(5,1)\times10^6, (10^7,10^6), (8\times10^6,5.333\times10^5), (8\times10^6, 4\times10^5)\}\,M_{\odot}$.  These mass combinations
correspond to mass ratios of $q = \{1.1, 5, 10, 15, 20\}$.  In all cases, the sources were placed at a redshift of $z=1$ and the time to
coalescence was set at $t_c=0.89$ yrs.  The input angular values were set as $\{\theta, \phi, \iota, \varphi_c, \psi\} =\{2.054,4.5,0.256, 3.707,1.794 \} $

\subsection{Setting a detection threshold}
Before investigating the detection possibilities of the GR waveforms,  we need to set a detection threshold for the eLISA observatory.  To do this, we 
conduct a null-signal test by assuming that the output of the 
detector is composed of instrumental noise only, i.e.\ $o(t) = n(t)$.  It is known that when a galaxy of white dwarf binaries is also
included in the data stream, an algorithm can be fooled into a false detection.  This is commonly known as the white-dwarf transform, where the
SMBHB signal is able to match power from the multitude of white dwarves at different frequencies and returns a $\rho > 0$.  While we do not include a galactic
foreground in this study, in the
same manner, it is also possible for a SMBHB template to match the random fluctuations of a Gaussian instrumental noise and
also return a positive SNR.

To set our threshold for detection, we ran fifteen algorithms, as described above, from different starting positions in the parameter
space.  In all cases, the algorithms returned ``detections" with SNRs of $9\leq\rho\leq 9.5$.  To account for the possibility of slightly higher values, we thus decided
to take $\rho=10$ as our threshold for detection.

\section{\label{Sec:Results}Results}
\subsection{Detection Horizons for non-GR theories}
Given the SNR threshold calculated in the previous section, our first objective is to calculate the detection horizons for the different
possible theories (i.e. the maximum redshift a system of a certain mass can be detected with $\rho \geq 10$).  To arrive at the various detection horizons, we use a Monte Carlo simulation based on an astrophysical distribution of sources~\cite{Sesana2010qb}.  For the Monte Carlo, we impose two restrictions : the first is that the maximum allowed redshifted
total mass corresponds to the MCMC limit of $m(z) = 1.163\times10^8\,M_{\odot}$.  The second is, for computational purposes,  to restrict the maximum array length for waveform generation to $2^{23}$ elements.  This second restriction will automatically exclude the investigation of some lower mass systems.  For each system, the angular values are
drawn from standard ranges, while the time of coalescence is chosen to be between $0.3\leq t_c/yrs\leq 0.99$.  Finally, $\beta$ is chosen from a uniform distribution
given the limits provided in Fig~\ref{Fig:upperlimits} for each value of $b$.

\begin{figure}[t]
 \includegraphics[width=\columnwidth,height=10cm]{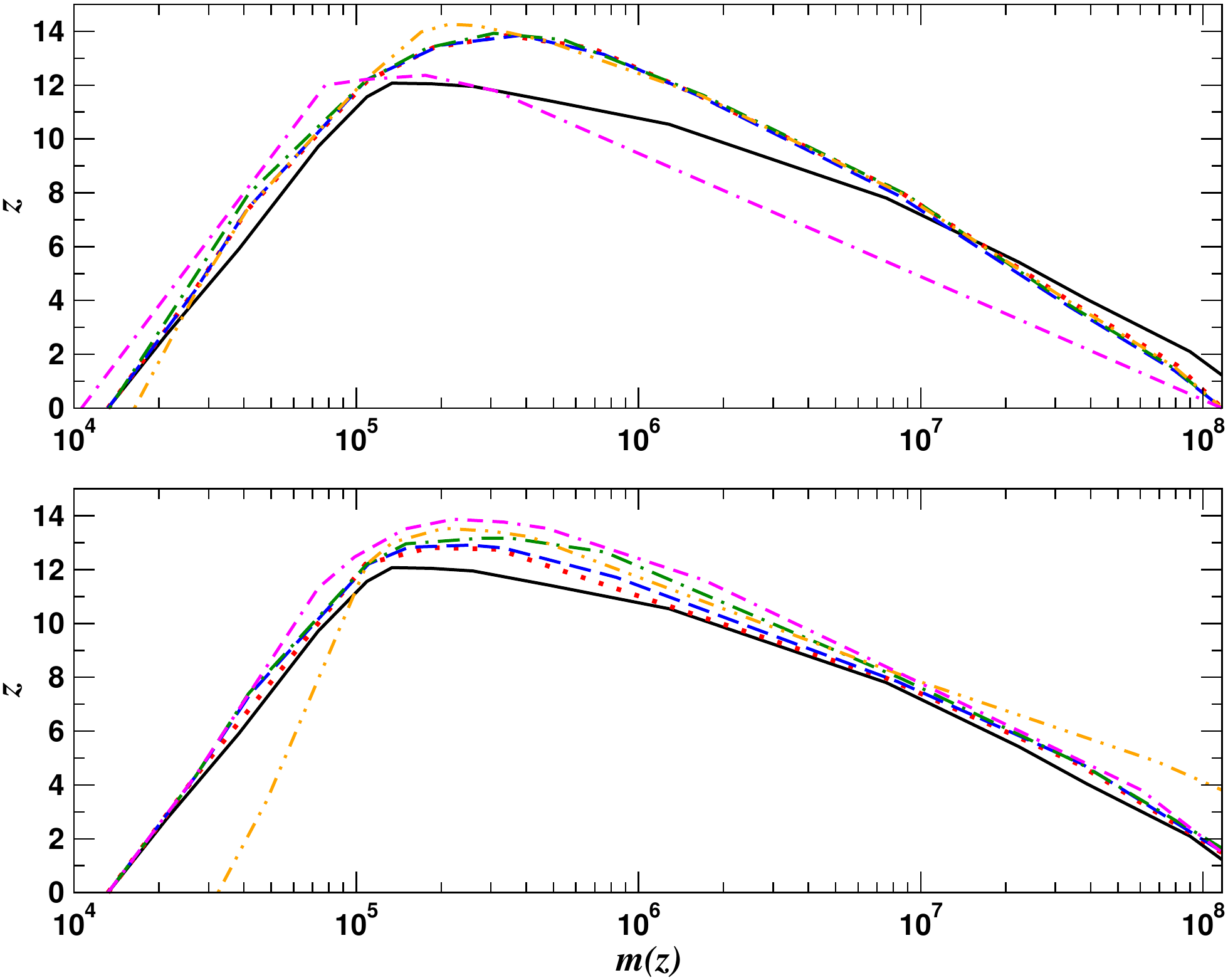}
 \caption{\label{Fig:horizons}Detection horizons as a function of redshifted total mass for positive (top figure) and negative (bottom figure) non-GR corrected
 waveforms.  In each figure we compare the different theories : GR (solid black line), $b=-1/3$ (red dotted line), $b=-2/3$ (blue dashed line), $b=-1$ (green dot-dashed line), $b=-4/3$ (orange double dot-dashed line), $b=-5/3$ (magenta double dash-dot line)}
\label{fig:horizons}
\end{figure}

In Fig~\ref{fig:horizons} we plot the detection horizons for both positive (upper panel) and negative (lower panel) corrections.  The various theories are represented
by : GR (solid black line), $b=-1/3$ (red dotted line), $b=-2/3$ (blue dashed line), $b=-1$ (green dot-dashed line), $b=-4/3$ (orange double dot-dashed line), $b=-5/3$ (magenta double dash-dot line).
If we assume that GR is the correct theory of gravity, we can see that for the eLISA detector, sources with redshifted total masses of
between $1.32\times10^4 \leq m(z)/M_{\odot} \leq 1.163\times10^8$ should be detectable with a SNR of $\rho \geq 10$.  For GR, the 
maximum of the redshift horizon peaks at $z\sim12$ for systems with $m(z)\sim10^5\,M_{\odot}$.

For positive corrections,  there is a marked difference between the $b=-5/3$ theory and all others.  Focusing first on the other theories,
we can see that at the high mass range, the redshift horizons for the alternative theories are shifted to lower total masses,  which is consistent with what we saw in Fig~\ref{fig:powerspectra}.  This is due to the fact that the dominant harmonics for these systems are
getting pushed to lower frequencies, thus reducing the detectability of the system.  Around the midrange of masses, we can see that the redshift horizons, and the peak of the horizon,  have increased to $z\sim14$ and  $m(z)\sim2-4\times10^5\,M_{\odot}$.  This increase
is due to the fact that these systems are having the dominant harmonics shifted into the minimum of the sensitivity curve given in 
Fig~\ref{fig:noise}.  As we approach the low mass end, there is a slight increase in the horizon over the GR value, but at masses of
less than $3\times10^4\,M_{\odot}$ they become equivalent to the GR horizon.  The main effect of the positive corrections can be seen
in the dominant $b=-5/3$ case.  Here, the redshift horizon is reduced by $\Delta z\sim2$ at high masses, and only crosses 
the GR horizon at a mass of $3\times10^5\,M_{\odot}$, which almost corresponds the point of peak sensitivity in the noise curve.  While
the maximum redshift is similar to the GR maximum, we can see that we do have a slight gain in the redshift horizon at low masses.

For negative systems, again there is one theory, $b=-4/3$, which has a different behaviour from all the rest.  This theory adds a 
correction at the 0.5PN order in phase and frequency.  At this order, the PN coefficients are equal to zero.  The addition of a negative
coefficient to the waveform at 0.5PN order clearly has a greater effect than the addition of a positive term at the same order.  In the
high mass range, the detection horizon is extended from $z\sim1$ to $z\sim4$.  This is due to the dominant harmonics of the waveforms
being moved to higher frequencies, allowing the detection of higher mass systems to higher redshift.  However, the price to be paid is
that we have a sharp drop off in the minimum masses detectable with $\rho\geq10$.  In the low mass range, as the frequencies of the
waveforms are being shifted to higher frequencies, the signals are being dominated by photon shot noise in the detector rendering
them undetectable.  For the alternative theories, we see an increase in the horizons as compared to the GR case, again extending from $z\sim12$ to $z\sim14$ depending on the theory, with a corresponding shift in the peak to higher masses.  At the low end, the horizons converge with the prediction from GR.

\subsection{Detection and Parameter Estimation of non-GR signals using GR templates}
The main goal of this work is twofold : firstly, given the pre-imposed maximum values of the coupling constant $\beta(b)$, what are the
limits of $\beta(q)$ where a non-GR signal, from a particular theory, can be detected with $\rho\geq10$ using a GR search template?
Secondly, given a positive detection, what are the limits of $\beta(q)$ such that the injected true binary parameters lie within a 
2$\sigma$ error bar of the recovered parameters as found by the MCMC?

\begin{figure*}
\begin{center}
\begin{minipage}{0.70\columnwidth}
\includegraphics[width=\textwidth]{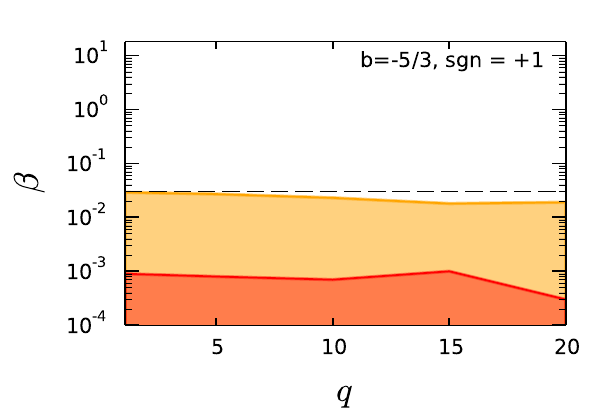}
\end{minipage}
\hspace{1cm}
\begin{minipage}{0.70\columnwidth}
\includegraphics[width=\textwidth]{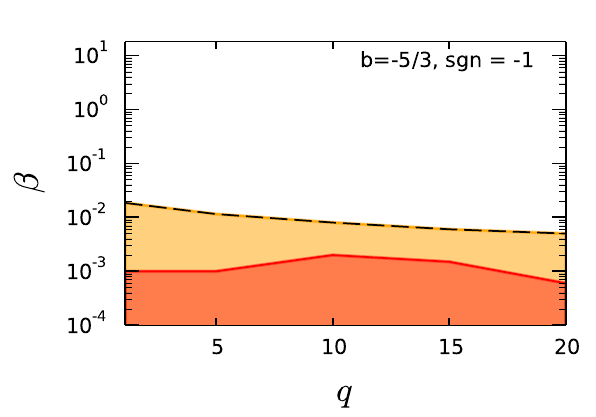}
\end{minipage}
\\
\begin{minipage}{0.70\columnwidth}
\includegraphics[width=\textwidth]{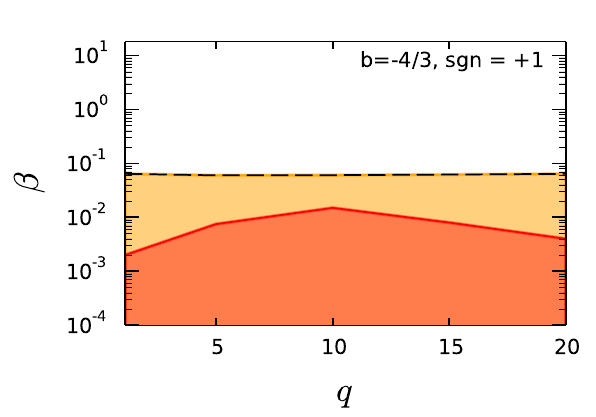}
\end{minipage}
\hspace{1cm}
\begin{minipage}{0.70\columnwidth}
\includegraphics[width=\textwidth]{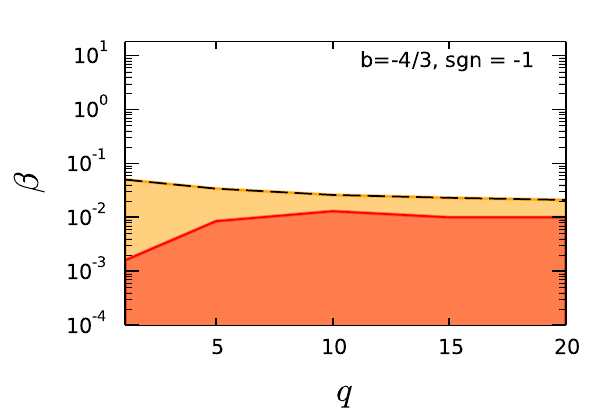}
\end{minipage}
\\
\begin{minipage}{0.70\columnwidth}
\includegraphics[width=\textwidth]{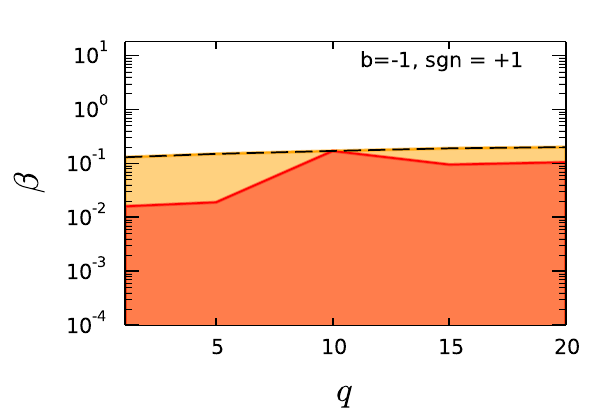}
\end{minipage}
\hspace{1cm}
\begin{minipage}{0.70\columnwidth}
\includegraphics[width=\textwidth]{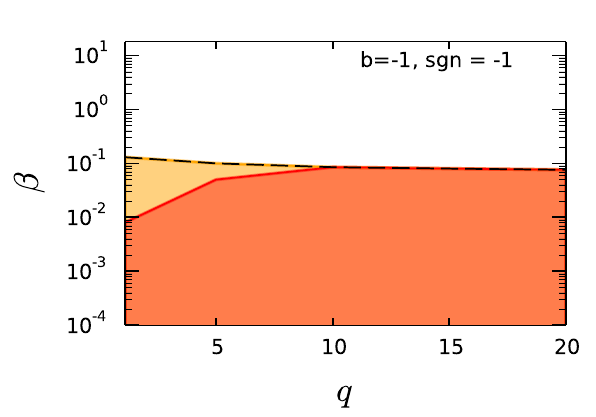}
\end{minipage}
\\
\begin{minipage}{0.70\columnwidth}
\includegraphics[width=\textwidth]{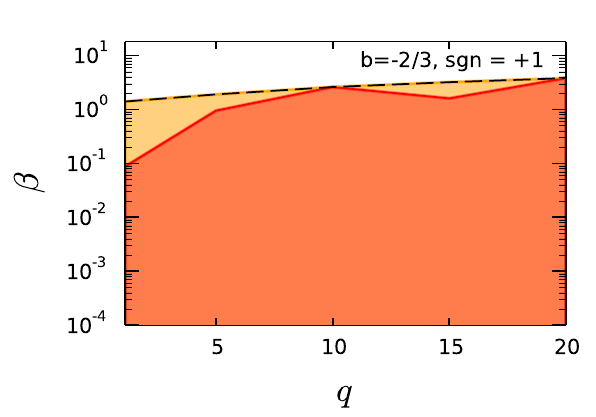}
\end{minipage}
\hspace{1cm}
\begin{minipage}{0.70\columnwidth}
\includegraphics[width=\textwidth]{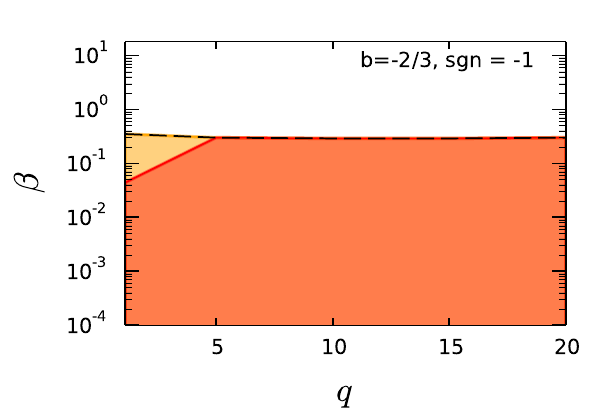}
\end{minipage}
\\
\begin{minipage}{0.70\columnwidth}
\includegraphics[width=\textwidth]{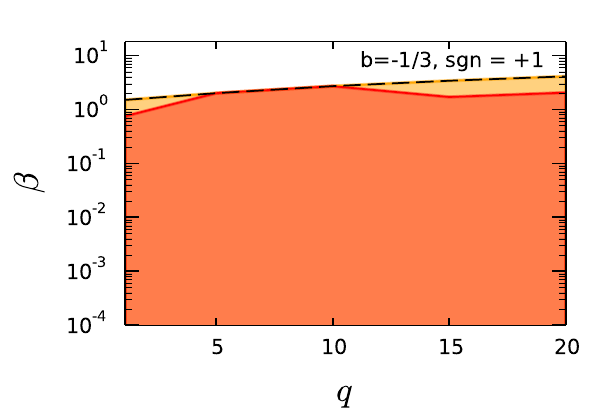}
\end{minipage}
\hspace{1cm}
\begin{minipage}{0.70\columnwidth}
\includegraphics[width=\textwidth]{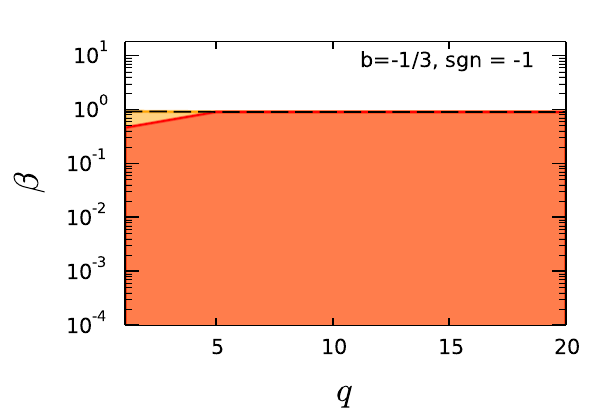}
\end{minipage}
\caption{Allowed upper limits for $\beta$ (dashed black line), detection limits (yellow curve) and limits where unbiased parameter estimation is still possible with GR templates (red curve).
We account for positive (left) and negative (right) corrections and corrections at increasing PN orders (i.e.\ increasing values of $b$ from top to bottom). 
Except for the positive correction with $b=-5/3$, GR templates manage to detect
all injected non-GR waveforms. With increasing values of $b$, the red curve below which GR templates are sufficient for parameter estimation approaches the detection limit.}
\label{Fig:Result}
\end{center}
\end{figure*}

We present the results of our analysis in Fig~\ref{Fig:Result}.  On the left of the image, we present results for the five different
alternative theories assuming positive corrections to the GR waveforms, while on the right, we display results assuming negative
corrections to the waveforms.  In each case, the maximum allowed value of $\beta$ is represented by the dashed black line, the limits
of detection by a GR template are given by the yellow shaded area, and the points at which a GR template is also sufficient for 
parameter estimation is represented by the red shaded area.

Let us first focus on the positive sign corrections : the panels on the left hand side demonstrate quite clearly that GR templates will be
sufficient for detection purposes in almost all situations.  The only region where non-GR templates are needed for detection is for the 
$b=-5/3$ theory when $q\geq8$.  For these mass ratios, and for values of $\beta\sim\beta_{max}$, the GR template fails to reach the
required SNR threshold.  In terms of parameter estimation, the GR templates perform better the higher the PN order at which the 
correction is introduced.  For the $b=-1$ to $b=-1/3$ theories, the vast majority of the parameter space is covered by GR templates.
Again it is only at $q\sim1$ and $q\geq10$ that non-GR templates are clearly needed for parameter estimation.  However, as we 
introduced non-GR corrections at lower PN orders, we see this situation change dramatically.  Now we observe that the size of the
yellow region growing, indicating that GR templates fail faster and faster.  In the dominant PN order correction theory ($b=-5/3$), we 
see that there is now a 1.5-2 order of magnitude range of $\beta$, for all values of $q$,  where non-GR templates are definitely needed
 for parameter estimation.
 
 In the negative correction case, we see a similar picture.  This time GR templates are sufficient for detection in all cases.  If we skip to
 the bottom right cell, for the $b=-1/3$ theory there is a small region of parameter space where, for values of $q\leq5$, non-GR templates
 are needed for parameter estimation.  As in the positive correction case, this region continues to grow the lower the PN order at 
 which the non-GR corrections appear, until again for the $b=-5/3$ theory, we have an entire band where for values of $\beta$ at all 
 values of $q$ GR templates are insufficient for parameter estimation.  This demonstrates that, regardless of the sign of the corrections,
 GR templates are pretty performant except in the cases where corrections are introduced at either the 0PN or 0.5PN order.
 
 \section{Conclusion}
 In this work, we have introduced a time domain waveform that can be used to test alternative theories of gravity, by introducing 
 leading order corrections at different PN orders in the phase and frequency of the waveform for comparable mass binaries.  The
 numerical FFT of this waveform has an accurate match to the Fourier domain SPA waveform that has been used for previous 
 studies within the ppE framework.  By construction, and with the intention of being able to conduct generic rather than specific alternative
 theory tests, the sign of the corrections in this non-GR waveform is arbitrary.   As solar system and binary pulsar tests have demonstrated
 no obvious deviations from GR, we imposed that the magnitude of the non-GR corrections should be no bigger than 50\% of the 
 corresponding PN coefficient.  
 
 After determining a SNR detection threshold of $\rho=10$ for a matched filtering algorithm using a model of the 
 future eLISA observatory, we conducted a full Bayesian 
 inference for multiple sources, given different values of the coupling constant $\beta$ and the mass ratio $q$.  We observed that in almost
 all cases, GR templates returned SNRs greater than the imposed detection threshold, making them more than adequate for the detection
 of signals, whether from GR or non-GR theories.  However, we determined that in the case of corrections appearing at the 0PN
 and 0.5PN orders in phase and frequency, non-GR templates were clearly needed to perform parameter estimation studies.  While the 
 performance of the GR templates improved the higher the PN order of correction, we still observed values of $\beta$, for $q\leq5$, where
 non-GR templates would also be necessary even in the 2PN order correction $b=-1/3$ theory.

This first study has allowed us to investigate the performance of GR templates in the detection and parameter estimation of non-GR signals
using observations of massive black hole binary signals with an eLISA GW observatory.  In future works, we intend to investigate the 
effect of possible changes to the eLISA mission configuration on tests of GR, and also to conduct a more intensive investigation into the regions of parameter space where GR templates look to be sufficient for both detection and parameter estimation.  A full Bayesian comparison is 
needed here to compare the full accuracy of both GR and non-GR templates when it comes to parameter estimation.

\begin{acknowledgments}
We would like to thank Nico Yunes for discussions and suggestions on the project.
C.H. is supported by the Swiss National Science Foundation.
\end{acknowledgments}

\clearpage



\bibliography{bibliography}

\end{document}